\documentstyle[12pt,prd,aps,epsfig,preprint]{revtex}

\begin{document}
\draft
\date{\today}
\title{Non-extensive Study of Rigid and Non-rigid Rotators}
\author{G\u{o}khan B. Ba\u{g}c\i$^{a}$, Ramazan Sever$^{a}$\thanks{%
Corresponding author: sever@metu.edu.tr}, Cevdet Tezcan$^{b}$}
\address{$^a$ Department of Physics, Middle East Technical University, 06531, Ankara,%
\\
Turkey \\
$^b$ Department of Electrical and Electronics Engineering, Baskent\\
University, 06530, Ankara, Turkey\\
}
\maketitle

\begin{abstract}
The isotropic rigid and non-rigid rotators in the framework of Tsallis
statistics are studied in the high and low \ temperature limits. The
generalized partition functions, internal energies and heat capacities are
calculated. Classical results of the Boltzmann-Gibbs statistics have been
recovered as non-extensivity parameter approaches to 1. It has also been
observed that non-extensivity parameter q behaves like a scale parameter in
the low temperature regime of the rigid rotator model.
\end{abstract}


\narrowtext
\newpage \setcounter{page}{1}

\section{\protect\bigskip Introduction}

\noindent \qquad The non-extensive generalization of the standard
Boltzmann-Gibbs \ statistics was proposed in 1988 by C. Tsallis[1-4]. This
non-extensive generalization begins with the supposition of a new, fractal
inspired entropy S$_{q}\equiv k(1-\sum_{i}p_{i}^{q})/(q-1)$ where q is any
real number and k is a positive constant which becomes Boltzmann constant in
the limit q$\rightarrow 1$. In this case, the new entropy S$_{q}$ takes the
form -k$_{B}\sum_{i}p_{i}\ln p_{i}$, which is simply the standard entropy
formula.This new non-extensive statistics has been studied a great deal to
list its properties and it is also applied to many well-known examples of
the Boltzmann-Gibbs statistics such as Ehrenfest theorem[5], Fokker- Planck
equations[6] and quantum statistics[7].

We first calculate the partition function of the isotropic rigid rotator for
both high energy and low energy limits and then obtain the generalized
internal energy and specific heat expressions for both cases. This forms the
Section II. The usual Maxwell-Boltzmann results are obtained in the limit q$%
\rightarrow 1$. In Section III, we turn our attention to non-rigid rotator
and follow the same procedure for it as in the Section II. Results and
discussions are given in Section IV.

\section{Isotropic Rigid Rotator}

Energy levels of a rigid rotator are[8]
\begin{equation}
E_{j}=j(j+1)\hbar ^{2}/2\mu a^{2}.
\end{equation}
where j=0,1,2, ..., $\mu =m_{1}m_{2}/(m_{1}+m_{2})$ reduced mass of the
nuclei and a being the equilibrium distance between them.The degeneracy of
each level is g$_{j}=2j+1.$ In non-extensive formalism (NEXT), the partition
function is given by
\begin{equation}
Z_{q}=\sum_{j=0}^{\infty }(2j+1)[1-\frac{\beta (1-q)j(j+1)\hbar ^{2}}{2\mu
a^{2}}]^{1/(1-q)}.
\end{equation}
Setting $\theta =\hbar ^{2}/2ma^{2}k,$ then we get
\begin{equation}
Z_{q}=\sum_{j=0}^{\infty }(2j+1)[1-\frac{(1-q)j(j+1)\theta }{T}]^{1/(1-q)}.
\end{equation}
Now we are ready to look for its analytic solutions in the high and low
temperature limits.

i)High temperature limit

At high temperatures, $\frac{\theta }{T}\ll 1$ and [1-$\frac{%
(1-q)j(j+1)\theta }{T}]^{1/(1-q)}$ changes slowly as j changes. So we take
it as a continuous function of j. Letting j(j+1)=x and substituting
(2j+1)dj=dx, then, we have
\begin{equation}
Z_{q}=\int_{0}^{\infty }dx[1-\frac{(1-q)x\theta }{T}]^{1/(1-q)}.
\end{equation}
For the interval 1 $\langle $ $q$ $\langle $ $2$, solution to the above
integral is
\begin{equation}
Z_{q}=\frac{T}{\theta }\frac{1}{2-q}.
\end{equation}
As q goes to 1, we get
\begin{equation}
Z_{q\rightarrow 1}=\frac{T}{\theta }.
\end{equation}
The expression above is exactly the Maxwell-Boltzmann (MB) partition
function for the isotropic rigid rotator in the high temperature limit.
Thus, we calculate the generalized internal energy function from
\begin{equation}
U_{q}=-\frac{\partial }{\partial \beta }\frac{Z_{q}^{1-q}-1}{1-q},
\end{equation}
in the non-extensive case. It becomes
\begin{equation}
U_{q}=\frac{1}{\beta ^{2}}[\frac{\alpha }{(2-q)}\beta ^{-1}]^{-q}\frac{%
\alpha }{(2-q)}~,~~~~~~~~1\langle q\langle 2
\end{equation}
where $\alpha \equiv \frac{1}{k\theta }$. It is important to see that again,
in the limit when q approaches to 1, we obtain the result already known in
MB statistics, i.e.,
\begin{equation}
U_{q\rightarrow 1}=\frac{1}{\beta }=k_{B}T.
\end{equation}
It is also possible to calculate the specific heat in this statistics as
\begin{equation}
C_{q}=\frac{\partial U_{q}}{\partial T},
\end{equation}
where U$_{q}$ is the internal energy function. We immediately get
\begin{equation}
C_{q}=(\frac{\alpha }{2-q})^{1-q}\frac{1}{k^{q-2}}(2-q)T^{1-q}.
\end{equation}
As q$\longrightarrow 1$, C$_{q}\longrightarrow k_{B}$ is verified easily.
All these calculations are carried out by using the constraint $%
\sum_{i=1}^{W}p_{i}^{q}\varepsilon _{i}=U_{q}$ . But, this choice of
internal energy constraint presents several difficulties. For example, the
probability distribution obtained by using this constraint is not invariant
under uniform translation of the energy spectrum. Another important
consequence of this constraint is that it leads to violation of energy
conservation macroscopically.The solution to all these difficulties are
recently proposed by C. Tsallis et al. [9]. They introduced the new internal
energy constraint as follows
\begin{equation}
\frac{\sum_{i=1}^{W}p_{i}^{q}\varepsilon _{i}}{\sum_{i=1}^{W}p_{i}^{q}}%
=U_{q}.
\end{equation}
This choice of constraint remedied all previous difficulties. Since we have
studied all thermodynamical quantities with the previous internal energy
constraint, we must reconsider them with this new constraint. There are two
ways to do this: Firstly, we can recalculate all thermodynamical quantities
with this new internal energy constraint by forming the new partition
function. Another method is to find the relation between the temperatu\i re
parameters of the old and new calculations. If one has all thermodynamical
functions calculated with the old constraint and the relation between
temperature parameters is known, it is possible to modify all previous
calculations carried out with the constraint $\sum_{i=1}^{W}p_{i}^{q}%
\varepsilon _{i}=U_{q}.$ This is the method we will follow, because we
already have the solutions with respect to old constraint.

We begin by writing all previous calculations in terms of intermediate
variable t$^{\prime }$ where t$^{\prime }\equiv 1/(\beta ^{\prime
}\varepsilon ).$ From now on, The superscript (2) will refer to calculations
done by old constraint.

The partition function of the rigid rotator in the high temperature limit
becomes
\begin{equation}
Z_{q}^{(2)}(\beta ^{\prime })=\frac{T^{\prime }}{\theta }\frac{1}{(2-q)}=%
\frac{t^{\prime }}{(2-q)}.
\end{equation}
Thus, we first evaluate the denominator in the Eq.(12)
\begin{equation}
\sum_{j}[p_{j}^{(2)}(\beta ^{\prime })]^{q}=\sum_{j}\frac{[1-(1-q)\beta
^{\prime }\varepsilon _{j}]^{q/(1-q)}}{Z_{q}^{(2)q}}.
\end{equation}
or it is explicitly written as
\begin{equation}
\sum_{j}[p_{j}^{(2)}(\beta ^{\prime })]^{q}=\frac{(2-q)^{q}}{t^{\prime q}}%
\sum_{j}[1-(1-q)\beta ^{\prime }\varepsilon _{j}]^{q/(1-q)}.
\end{equation}
By defining
\begin{equation}
\sum_{j}[1-(1-q)\beta ^{\prime }\varepsilon _{j}]^{q/(1-q)}=t^{\prime },
\end{equation}
we write it in the following simple form
\begin{equation}
\sum_{j}[p_{j}^{(2)}(\beta ^{\prime })]^{q}=(2-q)^{q}(t^{\prime })^{1-q}.
\end{equation}
We now proceed by using the mathematical relation between $\beta $ and $%
\beta ^{\prime }$ which is
\begin{equation}
\beta =\beta ^{\prime }\frac{\sum_{j}[p_{j}^{(2)}(\beta ^{\prime })]^{q}}{%
1-(1-q)\beta ^{\prime }U_{q}^{(2)}(\beta ^{\prime
})/\sum_{j}[p_{j}^{(2)}(\beta ^{\prime })]^{q}}.
\end{equation}
Substituting Eq. (17) into the Eq. (18), we get
\begin{equation}
t^{\prime }=(2-q)^{(q+1)/q}t^{1/q}.
\end{equation}
From Ref. [9], we have
\begin{equation}
Z_{q}^{^{\prime }(3)}(\beta )=Z_{q}^{(2)}(\beta ^{\prime }).
\end{equation}
By mere substitution, we get
\begin{equation}
Z_{q}^{^{\prime }(3)}(\beta )=(2-q)^{1/q}t^{-1/q}.
\end{equation}
We also have
\begin{equation}
p_{j}^{(3)}(\beta )=p_{j}^{(2)}(\beta ^{\prime }).
\end{equation}
Thus, we formed p$_{j}^{(3)}(\beta ).$ Next, we use the relation
\begin{equation}
\sum_{j}[p_{j}^{(3)}(\beta )]^{q}=\bar{(Z_{q}}^{(3)})^{1-q},
\end{equation}
to obtain the partition function
\begin{equation}
(\bar{Z_{q}}^{(3)})^{1-q}=(2-q)^{1/q}t^{(1-q)/q}.
\end{equation}
Therefore, one may obtain internal energy and heat capacities by using the
following relation
\begin{equation}
\beta \frac{\partial U_{q}^{(3)}}{\partial \beta }=\frac{\partial }{\partial
\beta }(\ln _{q}\bar{Z_{q}}^{(3)}).
\end{equation}
We find
\begin{equation}
U_{q}^{(3)}=(2-q)^{1/q}t^{1/q},
\end{equation}
where $ln_{q}x\equiv \frac{x^{1-q}-1}{1-q}$. Heat capacity is given by
\begin{equation}
C_{q}^{(3)}=\frac{dU_{q}^{(3)}}{dT}=\frac{k}{q}(2-q)^{1/q}t^{(1-q)/q}.
\end{equation}
Again, as q$\longrightarrow 1$, C$_{q}^{(3)}\longrightarrow k_{B}$ is
obtained. This is the rotational specific heat of the isotropic rigid
rotator for high temperature limit in accordance with the equipartition
theorem. Finally, we plot C$_{q}^{(3)}$ in terms of reduced temperature t=T/$%
\theta $ in Fig. 1.

ii) Low temperature limit

At low temperatures, if q is small enough concerning the non-extensive (1-q)
part, we have T$\ll \theta $. We look at first few terms of the summation in
the Eq. (3)
\begin{equation}
Z_{q}^{(2)}\cong 1+3[1-2(1-q)\frac{\theta }{T}]^{1/(1-q)}.
\end{equation}
This is nothing but the well-known partition function of the rigid rotator
in low temperature limit in MB statistics if (1-q) is small enough. This is
the case if we make q closer to 1. For such a choice, the partition function
takes the form
\begin{equation}
Z_{q\longrightarrow 1}^{(2)}\cong 1+3\exp (-2\theta /T).
\end{equation}
The generalized internal energy function \ for this case becomes
\begin{equation}
U_{q}^{(2)}=\frac{6}{\alpha }[1+3[1-2(1-q)\frac{\beta }{\alpha }%
]^{1/(1-q)}]^{-q}[1-2(1-q)\frac{\beta }{\alpha }]^{q/(1-q)}.
\end{equation}
This is a relatively long expression for the internal energy function but if
we look at the q$\rightarrow 1$ limit again, we see that it is of the form
\begin{equation}
U_{q\rightarrow 1}^{(2)}=\frac{6k_{B}\theta }{Z_{q\rightarrow 1}}\exp
(-2\theta /T),
\end{equation}
with 1/$\alpha =k_{B}\theta $ in the q$\rightarrow 1\lim $it$.$ Calculation
of the specific heat reads
\begin{eqnarray}
C_{q}^{(2)} &=&\frac{12q}{\alpha ^{2}kT^{2}}[1+3[1-2(1-q)\frac{\beta }{%
\alpha }]^{1/(1-q)}]^{-q}[1-2(1-q)\frac{\beta }{\alpha }]^{\frac{2q-1}{1-q}}
\nonumber \\
&&-\frac{36q}{\alpha ^{2}kT^{2}}[1+3[1-2(1-q)\frac{\beta }{\alpha }%
]^{1/(1-q)}]^{-q-1}[1-2(1-q)\frac{\beta }{\alpha }]^{\frac{2q}{1-q}}.
\end{eqnarray}
To modify the equations above in accordance with the third constraint, we
substitute the Eq. (19) into the Eq. (3) and take the first two terms to get
the low temperature limit partition function
\begin{equation}
Z_{q}^{(3)}\cong 1+3[1-2(1-q)(2-q)^{-(q+1)/q}\theta
^{1/q}T^{-1/q}]^{1/(1-q)}.
\end{equation}
Using this partition function, we immediately get the specific heat value as
\begin{eqnarray}
C_{q}^{(3)} &=&\frac{6k}{t^{1/q}}%
(2-q)^{-(q+1)/q}[1+3(1-2(1-q)(2-q)^{-(q+1)/q}t^{-q})^{1/(1-q)}]^{-q}
\nonumber  \label{e34} \\
&&[1-2(1-q)(2-q)^{-(q+1)/q}t^{-q}]^{q/(1-q)}.
\end{eqnarray}

As $q\longrightarrow 1$, it reduces to

\begin{equation}
C_{q}^{(3)}\longrightarrow \frac{12}{\alpha ^{2}k_{B}T^{2}}\exp (-2\theta
/T).
\end{equation}
This limiting value of heat capacity is obtained by observing that $\frac{1}{%
Z_{q\longrightarrow 1}^{(3)}}$ \ is almost equal to unity for low
temperature partition function values. This consideration also holds for Eq.
(33)

The plots related to low temperature case are illustrated again in terms of
the reduced temperature in Fig. 2. The interesting feature in these plots is
that specific heat function of the rigid rotator in the low temperature
regime attains the same shape as the classical one but with a narrower width
and a shift in the peak to the left. By increasing q, specific heat function
attains the same shape as the classical one but with a narrower width and a
shift in peak to the left.

\section{ISOTROPIC NON-RIGID ROTATOR}

In this case, energy levels are given by[10]
\begin{equation}
E_{j}^{(nonrigid)}=\frac{\hbar ^{2}}{2I_{0}}j(j+1)-\frac{\hbar ^{4}}{%
2I_{0}^{2}kR_{0}^{2}}j^{2}(j+1)^{2}.
\end{equation}
where $I_{0}=\mu R_{0}^{2}$ and k is spring constant. We have neglected the
third order term $j^{3}(j+1)^{3}$ in the above expression. It is better to
rewrite equation above in the simple form
\begin{equation}
\bar{E_{j}}^{(nonrigid)}=\frac{E_{j}^{(nonrigid)}}{hc}%
=Bj(j+1)-Dj^{2}(j+1)^{2}.
\end{equation}
where D is called centrifugal distortion constant. The first term on the
right hand side simply corresponds to the rigid rotator part which had been
studied in detail in the previous section, and non-rigidity term is
explicitly given as, apart from a multiplicative factor which is very small
compared to the rigidity part factor, $j^{2}(j+1)^{2}$. The partition
function in the NEXT formalism reads
\begin{equation}
Z_{q}^{(nonrigid)}=\sum_{j=0}^{\infty }(2j+1)[1-\beta (1-q)\bar{E_{j}}%
^{(nonrigid)}]^{1/(1-q)}.
\end{equation}
By substituting the related energy equation to the equation above, we get
\begin{equation}
Z_{q}^{(nonrigid)}=\sum_{j=0}^{\infty }(2j+1)[1-\beta (1-q)Bj(j+1)+\beta
(1-q)Dj^{2}(j+1)^{2}]^{1/(1-q)}.
\end{equation}

We again look for its solutions in the high and low temperature limits.

i)High temperature limit

We make the same assumptions used in the the rigid rotator case.
\begin{equation}
Z_{q}^{(nonrigid)}=\int_{0}^{\infty }dx[1-\beta (1-q)Bx+\beta
(1-q)Dx^{2}]^{1/(1-q)},
\end{equation}
or simply
\begin{equation}
Z_{q}=\int_{0}^{\infty }dx[ax^{2}+bx+1]^{1/(1-q)}.
\end{equation}
where
\begin{equation}
a=\beta (1-q)D~~~and~~~b=-\beta (1-q)B.
\end{equation}
The integral above can be rewritten in factorial form
\begin{equation}
Z_{q}=(mn)^{\frac{1}{1-q}}\int_{0}^{\infty }dx(x+\frac{1}{m})^{1/(1-q)}.(x+%
\frac{1}{n})^{1/(1-q)},
\end{equation}
where $m=b-n$ and $n=\frac{b}{2}\ [1\pm (1-4\frac{a}{b^{2}})^{1/2}]$. We
shall use the following general form to evalute the above integral
\begin{equation}
\int_{0}^{\infty }\ dxx^{\nu -1}(x+\beta )^{-\mu }(x+\gamma )^{-\sigma
}=\beta ^{-\mu }\gamma ^{\nu -\sigma }\ Beta(\nu ,\mu -\nu +\sigma
)_{2}F_{1}(\mu ,\upsilon ;\mu +\sigma ;1-\frac{\gamma }{\beta }),
\end{equation}
where $\mu =\frac{1}{q-1}$, $\nu =1$, $\gamma =\frac{1}{n}$ and $\beta =%
\frac{1}{m}$. ${_{2}F_{1}}(\rho ,\mu ;\gamma ;x)$ is the hypergeometric
function given below
\begin{equation}
_{2}F_{1}(\rho ,\mu ;\gamma ;x)=\frac{\Gamma (\gamma )}{\Gamma (\mu )\Gamma
(\gamma -\mu )}\int_{0}^{1}dtt^{\mu -1}(1-t)^{\gamma -\mu -1}(1-xt)^{-\rho },
\end{equation}

where the parameters satisfy $\nu >0$ and $\mu >\nu -\sigma $. Here we see
that $\nu =1$ and $q>1$. Thus, we simply get
\begin{equation}
\mu =\sigma =\frac{1}{q-1}.
\end{equation}
By using the condition $\mu >\nu -\sigma $, we obtain an upper limit for q
as $q<3$. Therefore, we have a solution range for q as $1<q<3$. Thus we get
the partition function as
\begin{equation}
Z_{q}=\frac{(q-1)}{(3-q)n}\text{ }_{2}F_{1}(1,1/(q-1);2/(q-1);1-\frac{m}{n}).
\end{equation}
. In order to get the compact expression for the partition function above,
we made use of the following identity:
\begin{equation}
Beta(1,\frac{2}{q-1}-1)=\frac{q-1}{3-q}.
\end{equation}

The resulting internal energy and heat capacity expressions may be found by
using Equations (7) and (10) respectively together with the partition
function above.Variation of heat capacity in the high temperature limit is
plotted in Fig. 3. In all the plots related to non-rigid rotator, we have
used 10.397 and 4.1$\times 10^{-5}$ as B and D parameters respectively for
HCl molecule [11].

ii) Low temperature limit

As in the case for isotropic rigid rotator in the low temperature limit, we
look at first few terms of the summation in the Eq. (39)

\begin{equation}
Z_{q}\cong 1+3[1-2B(1-q)\beta +4D(1-q)\beta ]^{\frac{1}{1-q}}.
\end{equation}

Internal energy term resulting from the partition function above is
\begin{equation}
U_{q}=6(B-2D)[1+2(B-2D)(q-1)\beta ]^{\frac{q}{1-q}}[1+3(1+2(B-2D)(q-1)\beta
)^{\frac{1}{1-q}}]^{-q}.  \nonumber
\end{equation}
The specific heat function of non-rigid rotator for this case is illustrated
in Fig. 4, with using B and D constants for HCl molecule, 10.397 and 4.1$%
\times 10^{-5}$ again.

\section{RESULTS AND DISCUSSIONS}

We have studied isotropic rigid and non-rigid rotators in the non-extensive
statistics by extending the previous work on anisotropic rigid rotator[12].
We have calculated heat capacity at high and low temperature limits for each
rotator. At high temperature, results are in well agreement with classical
cases. We also observe some interesting features for low temperature case.
There happens to be shift to the left in the zeros of internal energy
functions by increasing q as is plotted in Fig. 2. By increasing q from 1 to
3/2, specific heat function preserves the same shape as the q=1 case but
with a narrower width and a shift to the left. One of the interesting
features of Tsallis statistics is that in the limiting Tsallis index, i.e.
as q$\rightarrow 1$, one gets the corresponding Maxwell-Boltzmann result at
once. This has also been true for the partition functions, generalised
energy functions and specific heats in the high and low temperature limits
of the isotropic rigid rotator. We have concluded that study of extensive
systems through non-extensive thermostatistics bears importance[13].

\section{ACKNOWLEDGEMENTS}

We thank S. Abe for his helpful comments. One of us (G. B. B.)acknowledges
C. Tsallis for creative environment in the International School and Workshop
on Non Extensive Thermodynamics and Physical Applications held in \
Villasimius/Italy in May 23-30 2001.

\newpage

\begin{figure}

\caption{Specific heat of the rigid rotator as a function of
reduced temperature t=T/$\theta $ in the high temperature limit.}

\par

\begin{picture}(161,265)
\put(-10,200){\psfig{file=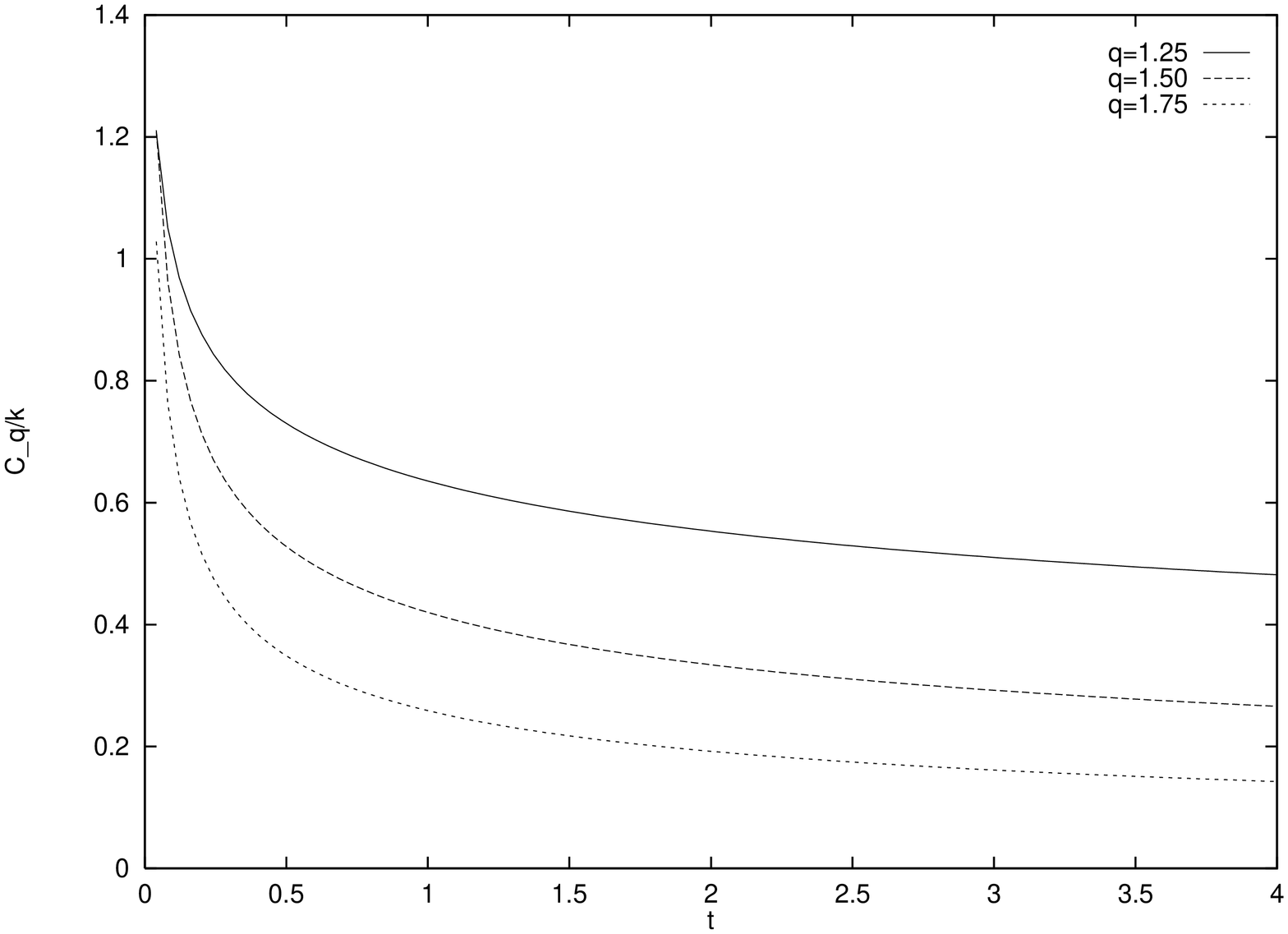,width=12cm,angle=270}}

\end{picture}
\label{Fig1}
\end{figure}

\newpage

\begin{figure}

\caption{Specific of the rigid rotator heat as a function of
reduced temperature t=T/$\theta $ in the low temperature limit.}

\par

\begin{picture}(161,265)
\put(-10 ,155){\psfig{file=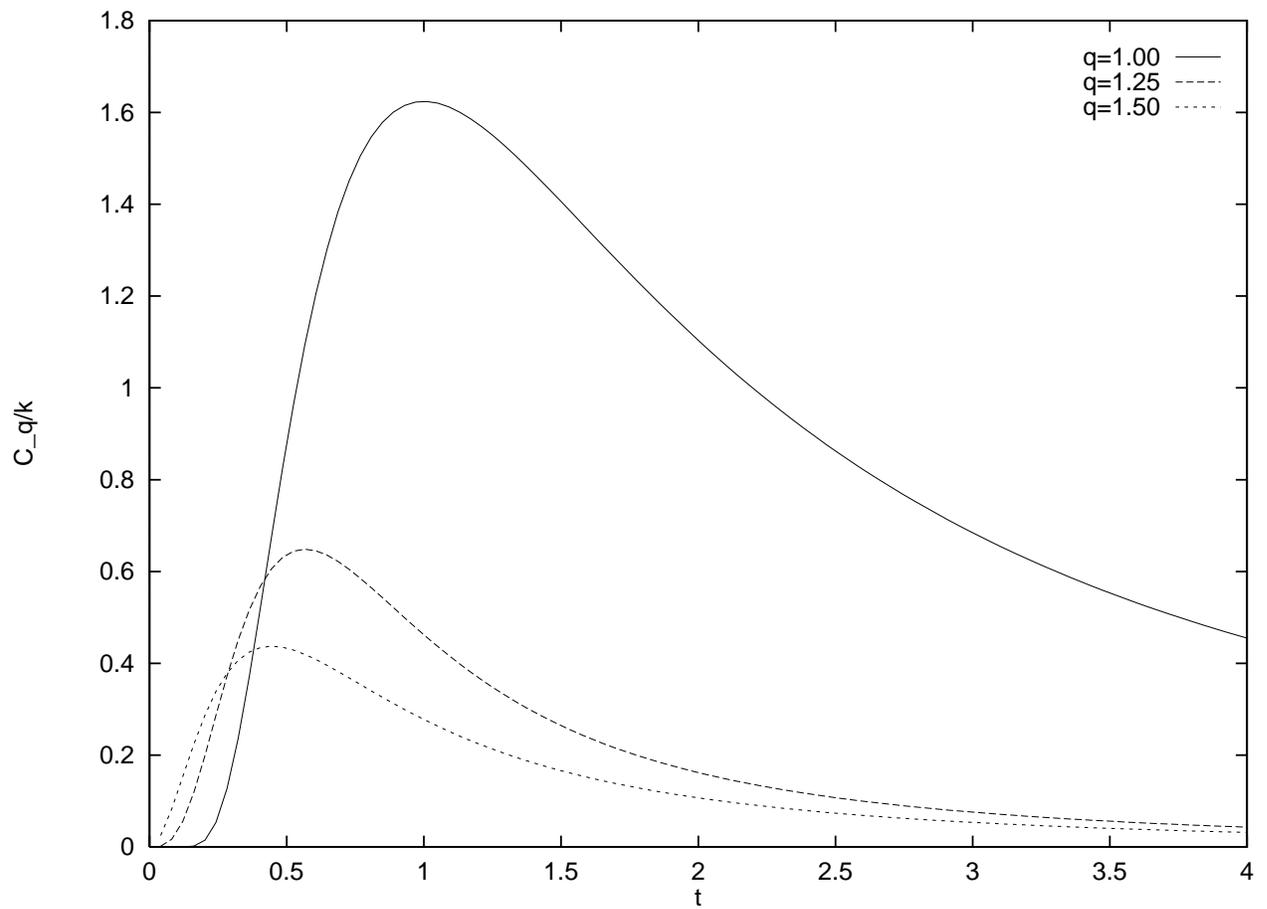,width=12cm,angle=270}}

\end{picture}
\label{Fig2}
\end{figure}

\newpage

\begin{figure}

\caption{Specific heat of the non-rigid rotator as a function of
temperature in the high temperature limit.}

\par

\begin{picture}(161,265)
\put(-5,-30){\psfig{file=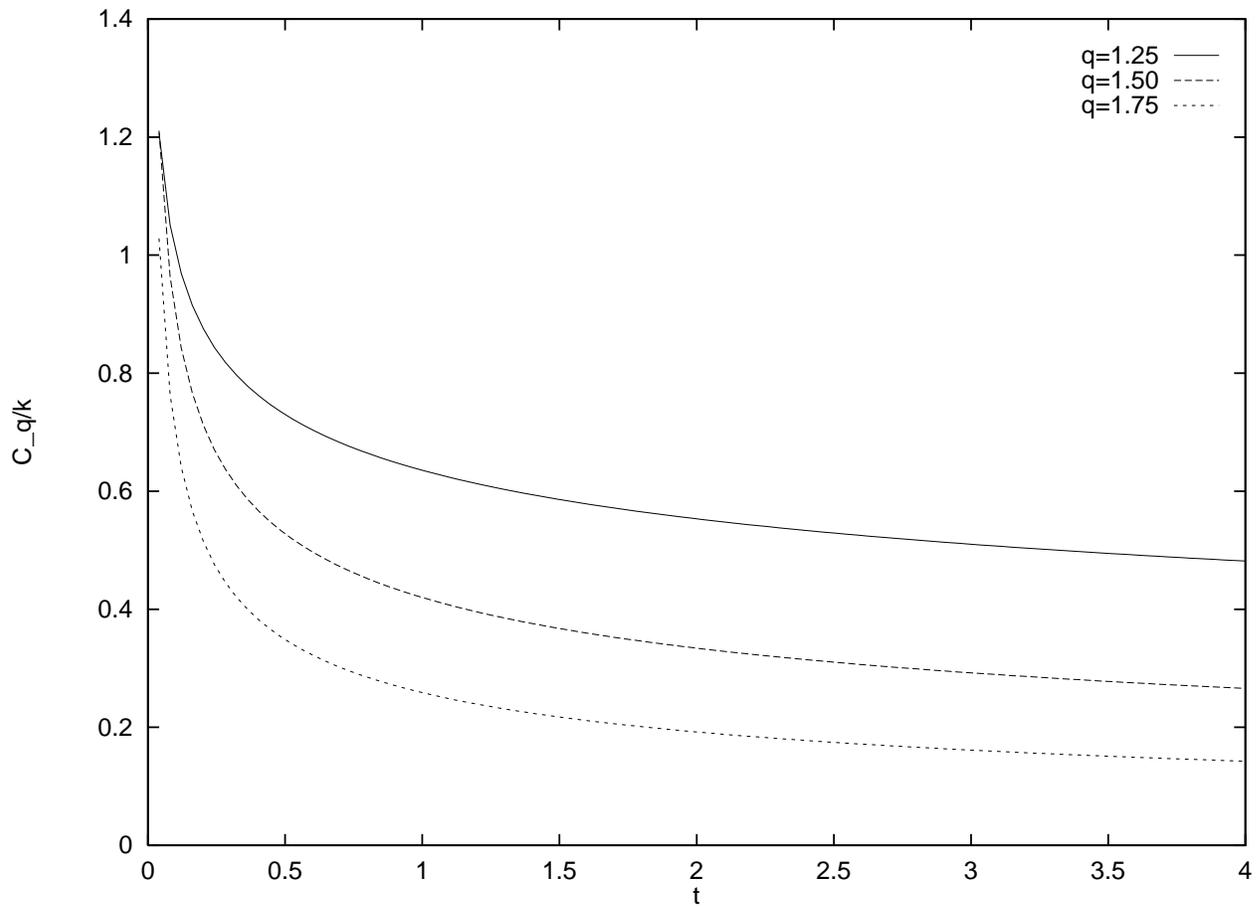,width=12cm,angle=270}}
\end{picture}
\label{Fig3}

\end{figure}

\newpage

\begin{figure}

\caption{Specific heat of the non-rigid rotator as a function of
temperature in the low temperature limit for q= 1.25, 1.50 and
1.75 respectively starting from above.}

\par

\begin{picture}(161,265)
\put(-10,-280){\psfig{file=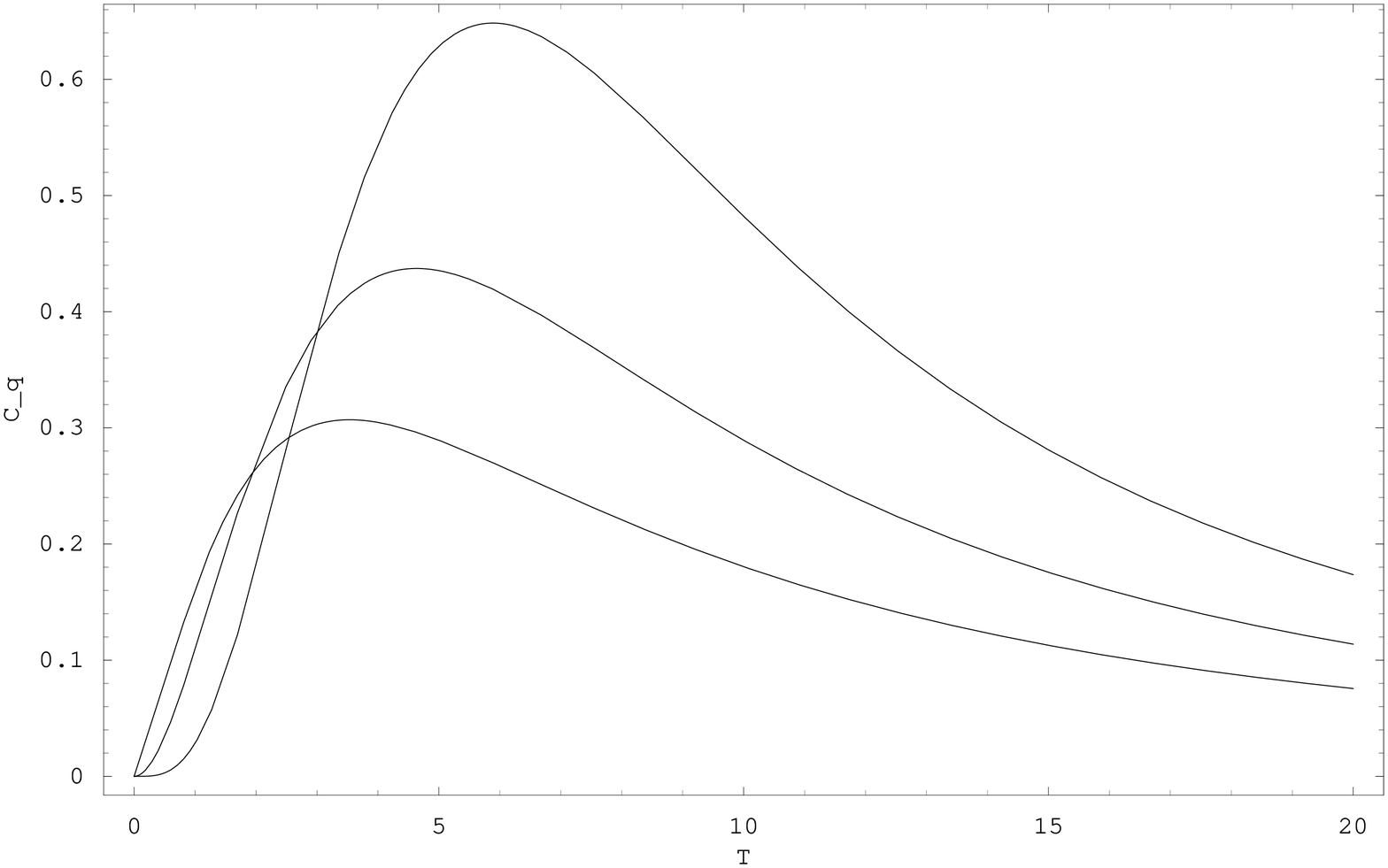,width=16cm,height=18cm}}

\end{picture}
 \label{Fig4}
\end{figure}

\end{document}